\documentclass [11pt] {article}

\begin{document}

\title{%
Observer-independent quantum of mass }
\author{ J.\ Kowalski--Glikman\thanks{e-mail
address jurekk@ift.uni.wroc.pl}\\ Institute for Theoretical
Physics\\ University of Wroc\l{}aw\\ Pl.\ Maxa Borna 9\\
Pl--50-204 Wroc\l{}aw, Poland} \maketitle

\begin{abstract}

It has been observed recently by Giovanni Amelino-Camelia
\cite{gac1, gac2} that the hypothesis of existence of a minimal observer-independent
(Planck) length scale is hard to reconcile with special relativity. As a remedy
he postulated to modify special relativity by introducing an
observer-independent length scale. In this letter  we set forward
a proposal how one should modify the principles of special
relativity, so as to assure that the value of mass 
scale is the same for any inertial observer. It turns out that
one can achieve this by taking dispersion relations such that the
speed of light goes to infinity for finite momentum (but infinite
energy), proposed  in the framework of the quantum
$\kappa$-Poincar\'{e} symmetry. It follows that at the Planck scale
the world may be non-relativistic.

\end{abstract}

\clearpage

In the recent years we face a growing mass of evidence (see e.g., 
\cite{gara}), coming both from loop quantum gravity, where one finds that 
area and volume are quantized \cite{rovsmol}, and from many aspects of  
string theory that space is quantized, i.e., there exists in nature a 
minimal length, usually identified with the Planck length ${\cal L}_P$. 
However, as pointed out recently by Giovanni Amelino-Camelia in a series of 
remarkable papers \cite{gac1}, \cite{gac2} the hypothesis of existence of 
the fundamental minimal length is by itself puzzling. If there is something 
fundamental about the Planck length  (i.e., if it has a status similar to 
that of the speed of light in special relativity), then it must have the 
same value for all inertial observers, which is hard to reconcile with one 
of the most basic results of special relativity, the FitzGerald-Lorentz 
contraction. So if we believe in the modern evidences, there are two 
choices: either to assume that the existence of the length scale reflects a 
property of some background field configuration which furnishes, what we 
call our universe\footnote{And thus the emergence of the scale has a 
dynamical origin.}, or, following \cite{gac1}, \cite{gac2}, to assume that 
the existence of the scale reflects  fundamental, kinematical properties of 
space-time. In the latter case it follow from the relativity postulate (see 
below) that one should assume that that the fundamental scale has to be the 
same for all inertial observers. If we make such assumption, there is no 
choise, but to modify the principles of special relativity. Such a 
modification has been proposed in \cite{gac1}, \cite{gac2}.   In these 
papers the author proposes to promote a minimal length to the status of the 
speed of light in the standard special relativity, i.e., to assume that the 
value of the minimal length is observer-independent. To do so 
Amelino-Camelia proposes to consider a theory with non-standard dispersion 
relation for light (and thus with variable speed of light) and to 
illustrate this proposal he presents some simple, leading order 
computations. His proposal is a starting point of our analysis presented 
below.

Before turning to our investigations, let us make the following
observation. There are three dimensionful constants in fundamental
physics: speed of light $c$ (as it will turn out this constant is
only a long-wavelength limit of velocities of massless particles),
Newton's gravitational constant $G$, and the Planck constant
$\hbar$. All these constants should play a fundamental role in the
quantum theory of gravity. Putting another way, in
 physics we have three fundamental scales, of length
 ${\cal L}_P = \sqrt{\hbar G/c^3} \sim 10^{-35}\, m$,
 time ${\cal T}_P = \sqrt{\hbar G/c^5} \sim 10^{-43}\, s$, and mass
 ${\cal M}_P = \sqrt{\hbar /G c} \sim 10^{-8}\, kg$.  Now, the problem
 is that if we believe in what special relativity teaches us,
 these scales are not observer-independent. We encounter therefore a paradox:
 On the one
 hand one would like the fundamental scales
  behave very much like the speed of light behaves in
 special relativity (and this is the heart of Amelino-Camelia's observation),
  i.e., if any inertial observer attempts to
 measure them, he/she gets the same result, and on the other, any
 Lorentz-boosted observer
 would attribute to them different values.

In this paper, instead of considering the minimal length we will 
concentrate on a related problem of 
 maximal mass, 
and only briefly comment on the minimal length issue.

 Our starting point would be  a set of the following postulates, being
 an slightly modified version of the postulates presented in \cite{gac1}

 \begin{enumerate}

 \item ({\em Relativity principle}) The laws
 of physics take the same form in all inertial frames.
\item ({\em Speed of light}) The laws of physics
involve the fundamental velocity scale $c$. This scale can be measured by
each inertial observer as a speed of light with wavelength much longer than 
the fundamental length $\lambda {\cal L}_P$. The speed of light depends on 
the wavelength $\lambda$ in such a way that it becomes infinity for finite 
$\lambda$ (of order of Planck length ${\cal L}_P$.) 
\item ({\em Mass scale}) The laws of physics
involve fundamental mass  scale, ${\cal M}_P$  which is the same
for all inertial observers. This mass scale is related to the
length scale as follows. For a photon of momentum ${\cal M}_P c$,
the wavelength $\lambda ={\cal L}_P$.
 \end{enumerate}

Let us observe that the first part of the second postulate is much
weaker than the analogous postulate in Einstein special
relativity, where it is assumed that the speed of light does not
depend on the wavelength and thus defines a universal tool for
measuring space-time distances between events. On the other hand
we know from the quantum theory that the wavy nature of light (and
all matter) has fundamental character, and one cannot avoid taking
it into account while considering a theory which is supposed to
describe the quantum nature of space and time.
\newline

It is clear from the second postulate that our starting point to
modify special relativity would be to allow for deviation from the
standard dispersion relation for photons so as to allow for
variable speed of light:

\begin{equation}\label{1}
 E^2 - c^2 p^2 = 0
\end{equation}
is to be raplaced by
\begin{equation}\label{2}
  {\cal F}\left(E, p; c, {\cal L}_P, {\cal M}_P\right) =0.
\end{equation}

Solving this equation for $E = f\left(p; c, {\cal L}_P, {\cal
M}_P\right)$ we can define the variable speed of light ${\cal C}$
to be
\begin{equation}\label{3}
{\cal C} = \frac{\partial E}{\partial p}.
\end{equation}

Now it is easy to see what we need in order to satisfy our postulates. In 
the standard special relativity, one finds that masses and distances are 
observer-independent in non-relativistic limit, i.e., when $V/c 
\rightarrow0$. We would encounter the same effect if we assume that the 
variable speed of light (\ref{3}) goes to infinity for some finite value of 
momentum carried by the light wave, that is, for some finite value of its 
wavelength. This means that the modified relativity has two Galilean 
limits: one in the non-relativistic limit $V/c \ll 1$ and $\lambda/{\cal L}_P 
\gg 1$ and the second in the Planck regime $\lambda/{\cal L}_P \sim 
1$\footnote{It i clear that if a theory being an extension of special 
relativity predicts ${\cal C} \rightarrow \infty$ in the ultra high energy 
regime, this regime should be Galilean. The reason is that this theory, to 
be consistent, must have special relativity as its limit, which in turn 
reduces to Galilean physics in the limit $V/c \ll 1$. But then both limits 
are to be equivalent.}.

 In what follows we will be
interested in a particular form of ${\cal F}$ which arises in the so-called 
quantum $\kappa$-Poincar\'{e} theory \cite{lunoruto, maru, luruto, luruza}.
This theory  results from applying the ideas of quantum deformations to 
four-dimensional Poincare algebra, and leads to modifications of 
relativistic symmetries at the energy scales comparable to the $\kappa$ 
parameter of the theory, which we will identify with ${\cal M}_P c$.

 Among different
realizations of quantum $\kappa$-Poincar\'{e}, we will be particularly
interested in the so called $+$-bicrossproduct basis \cite{maru, 
luruza}\footnote{This realization has a virtue that the Lorentz sector, as
well as the action of rotations on momenta are undeformed.}, in which 
(restricted to two dimension) infinitesimal action of boosts $N$, with 
parameter $\omega$ takes the form 
\begin{equation}\label{a}
 \delta p = \omega\,\left[\frac{{\cal M}_P c}{2} \left( 1 - e^{-2 E/{\cal M}_P
 c^2}\right) - \frac{1}{2{\cal M}_P c} p^2\right]
\end{equation}

\begin{equation}\label{b}
\delta E = \omega\, c p
\end{equation}

One can easily check that the following dispersion relation  is invariant 
under these transformation rules (in the case of massive particles, one 
should replace $0$ on  the right hand side with $m^2 c^4$), i.e., the 
expression below is a Casimir of the $\kappa$-Poincar\'{e} algebra: 
\begin{equation}\label{4}
 \left(2\,{\cal M}_P c^2\,\sinh \frac{E}{2{\cal M}_P
 c^2}\right)^2\,
  -c^2p^2 e^{E/{\cal M}_Pc^2} =0.
\end{equation}

 Of course, in the
 limit of large  ${\cal M}_P$, i.e., $E/{\cal M}_P c^2 \ll 1$,
 $p/{\cal M}_P c \ll 1$, from (\ref{a}), (\ref{b}), and (\ref{4}) one obtains the
 standard boost action and the dispersion relation (\ref{1}), respectively.

 The dispersion relation (\ref{4}) has a remarkable property, that it
 furnishes a theory
  obeying the postulates presented above. Indeed if we write it in the form
\begin{equation}\label{5}
 {\cal M}^2_P c^4\,\left(1- e^{-E/{\cal M}_P
 c^2}\right)^2\,
  -c^2p^2 =0,
\end{equation}
  it is easy to see that when $E\rightarrow\infty$,
  $p\rightarrow {\cal M}_P c$, i.e., the energy of the wave with
  finite length is infinite. Putting differently, the wavelength dependent
  speed of
  light
\begin{equation}\label{6}
 {\cal C}(p) = \frac{dE}{dp} = c \left(1 - \frac{p}{{\cal M}_P c}\right)^{-1} = c\exp(E/{\cal M}_P c^2)
\end{equation}
tends to infinity when $p\rightarrow {\cal M}_P c$. It should be stressed 
that in order to make the speed of light infinite one should use an 
infinite amount of energy, similarly to the standard special relativistic 
case, when  one wants to make a massive particle to move with the speed of 
light.

Now it is easy to see that it follows from the infinitesimal
transformations (\ref{a}, \ref{b}) that all inertial observers
would measure the same value of ${\cal M}_P$. Indeed $$ \delta
p|_{p={\cal M}_P c} = \lim_{E \rightarrow \infty} \left.
\omega\left[\frac{{\cal M}_P c}{2} \left( 1 - e^{-2 E/{\cal M}_P
 c^2}\right) - \frac{1}{2{\cal M}_P c} p^2\right]\right|_{p={\cal M}_P
 c}=0.$$ In this way we satisfy third
 postulate. This is not very surprising after all, because in the limit 
$p\rightarrow {\cal M}_P c$ the speed of light becomes infinite, and the 
theory becomes effectively Galilean.

Let us observe   now that in order to measure distances of the length 
$\ell$ we need to have in our disposal a photon of the length $\lambda \sim 
\ell$.  Let us assume moreover that, like at low energies, momentum of the wave 
is inverse proportional to the wavelength $p \sim 1/\lambda$. Then simple 
dimensional analysis leads us to the  expression $p \sim {\cal M}_P c \, 
{\cal L}_P/\lambda$, where ${\cal L}_P$ is the Planck  length. But this 
means that if the relation (\ref{4}) holds there must exist a minimal 
observable length equal exactly ${\cal L}_P$ (up to a numerical factor.) 
The same conclusion is true of course if one replaces (\ref{4}) with any 
other dispersion relation with the property that energy goes to infinity 
for finite value of momentum. Thus the theory leads to predicting the 
existence of the minimal length. What we need to check is if this length 
would be the same, when measured by any inertial observer. Since in the 
relevant limit the theory is Galilean, it is almost obvious that this must 
be a case. However in order to prove this result one must extend the theory 
from the momentum sector to the whole phase space, which is non-commuting 
in the position sector (reflecting in this way a quantum character of the 
$\kappa$-Poincare group.) This will be done in a separate paper. 
\newline

Let us complete this letter with a number of comments.
\begin{enumerate}
\item In this paper we worked in the two-dimensional framework. In
$D=4$ transformation (\ref{a}) takes the form $$\delta p_i =
\omega_{i}\frac{{\cal M}_P c}{2} \left( 1 - e^{-2 E/{\cal M}_P
 c^2}\right) +  \frac{1}{2{\cal M}_P c}\left(\delta_{ij} p^2
 - \frac12 p_ip_j\right)\omega^j$$ It should be noted that the second term
 is a conformal boost, so that the transform is a sum of a
 (deformed) standard boost and the conformal one. The $4D$
 transformations and their physical implications will be
 investigated in a separate paper \cite{RGJ}.
\item One should observe that the theory presented here is  fully
falsifable even at this very premature stage. First of all it is fully 
consistent with all present experimental data  \cite{grbgac}. Second there 
are proposal of experiments to be performed in the coming years, aimed at 
checking the cnsequences of the moodified dispersion relations of the form 
 (\ref{4}) \cite{glast}.
\item The result of this paper, namely the prediction concerning the 
energy dependence of the speed of light might find its application in the 
cosmological models in which the variable speed of light makes it possible 
to solve the well known problems of the standard cosmological model 
\cite{cosmo}. It should be noted that in the model presented here the speed  
of light grows with energy, so it should be much higher than $c$ in the 
very early universe, the behavior which would  in principle allow to 
resolve cosmological puzzles. 
\item Last, but not least, it should be observed that if the main
result of this letter is correct, namely that the existence of the 
observer-independent fundamental mass  scale results from the fact that the 
velocity of light goes to infinity for finite wavelength, it follows that 
physics on Planck scale is not governed by any relativistic theory. Rather, 
the theory of space, time and processes at this scale, i.e., the theory of 
quantum gravity should be a Galilean theory possibly (given the 
non-commutative space-time structure resulted from the quantum algebra 
structure of the $\kappa$-Poincar\'{e} algebra \cite{maru}, \cite{luruza}) with 
a discrete, non-commutative space and  time. This idea  in the author's 
opinion certainly deserves further investigations. 
\end{enumerate}

{\bf Acknowledgements}. I would like to thank Giovanni Amelino-Camelia for 
his constant encouragement, and stimulating comments. I would never  start 
thinking about the problems addressed in this paper had it not for his 
passionate explanation of his results presented in \cite{gac1, gac2}. I 
have also benefited a lot from discussions with Jerzy Lukierski and Arek 
B\l{}aut. This work was partially supported by KBN grant 5PO3B05620.

\end{document}